\documentclass[11pt]{article}
\usepackage{amssymb,amsmath,amsthm,amscd,latexsym}
\usepackage{mathrsfs}
\usepackage{mathrsfs}
\usepackage{amsfonts}
\usepackage{amsmath}
\usepackage{amssymb}
\usepackage{amscd}
\usepackage{graphicx}
\usepackage{epstopdf}

\renewcommand{\paragraph}{\roman{paragraph}}
\input cyracc.def

\newcommand{\F}{\mathbb{F}}

\begin{document}
\title{\bf Optimal three-weight cubic codes
\thanks{This research is supported by National Natural Science Foundation of China (61672036), Technology Foundation for Selected Overseas Chinese Scholar, Ministry of Personnel of China (05015133) and the Open Research Fund of National Mobile Communications Research Laboratory, Southeast University (2015D11) and Key projects of support program for outstanding young talents in Colleges and Universities (gxyqZD2016008).}
}
\author{\small{Minjia Shi
\thanks{Corresponding author.}}\\ \small{Key laboratory of Intelligent Computing \& Signal Processing, Ministry of Education, Anhui}\\
\small{ University No. 3 Feixi Road, Hefei Anhui Province 230039, P. R. China, National Mobile}\\
\small{Communications Research Laboratory, Southeast  University and School of Mathematical}\\
\small{Sciences of Anhui University, Anhui, 230601, P. R. China}
\and\small{Hongwei Zhu}\\ \small{School of Mathematical Sciences, Anhui University, Hefei, Anhui, 230601, P. R. China}
\and\small{Patrick Sol\'e}\\ \small{CNRS/ LAGA, University Paris 8, 93 526 Saint-Denis, France}\\
}
\date{}
\maketitle
{\bf Abstract:} {In this paper, we construct an infinite family of three-weight binary codes from linear codes over the ring
$R=\mathbb{F}_2+v\mathbb{F}_2+v^2\mathbb{F}_2$, where $v^3=1.$ These codes are defined as trace codes. They have the algebraic structure of abelian codes.
Their Lee weight distributions are computed by employing character sums. The three-weight binary linear codes which we construct are shown to be optimal when $m$ is odd and $m>1$.
They are cubic, that is to say quasi-cyclic of co-index three.
An application to secret sharing schemes is given.

\section{Introduction}
\hspace*{0.5cm} Since the weight distribution of a code is a crucial tool to estimate the error correcting capability, and the probability of error detection and correction,
it is a fundamental topic in Coding Theory, and has been studied in many papers \cite{DD,DY,DZ,HY,SL,SS}.
Linear codes with a few weights received a special attention, because these codes enjoy interplay with strongly regular graphs,
partial geometries, and have some applications in secret sharing schemes, such as \cite{AB,AE,RWM,DP,DD,YC,SL,YJ}. In particular, three-weight codes over finite fields
have been studied since the 1980s, due to their connections with association schemes and finite geometries \cite{CG,RWM}. Note that many very interesting codes
(see \cite{DY,DZ,HY}) can be produced by considering trace codes over finite fields. Later,
in \cite{SL, SP, SR}, the authors have constructed several infinite family of a few weights codes from trace codes over different rings.


In this paper, we employ an evaluation map, and a linear Gray map from a cubic ring extension $R$ of $\F_2$ to  $\F_2^3$ to define our codes. This ring appears to characterize quasi-cyclic codes
of co-index three, sometimes called {\em cubic codes} \cite{B,LS}. While our construction of binary codes, by the choice of the Gray map is not exactly the classical cubic construction
of \cite{LS}, the binary codes we construct are still cubic, as it is easy to see. They  are provably abelian easily
but perhaps not cyclic.
While most constructions of two-weight codes, or three-weight codes in the literature are based on cyclicity and cyclotomy \cite{AE},
our codes are defined as trace codes. Their coordinate places are indexed by the group of units of an algebraic extension of a finite ring.
Their weight distributions are determined by using exponential character sums and the Chinese Remainder Theorem.
Taking a linear Gray map, we obtain an infinite family of binary three-weight codes.
The latter codes are shown to be optimal for given length and dimension by the application of the Griesmer bound when they cater to certain conditions.\\
\hspace*{0.5cm}The manuscript is organized as follows. Basic notations and definitions are provided in Section 2. Section 3 lists the main results of this paper. The proofs of Theorems 3.1, 3.2, 3.3 and 3.4 are given in Section 4. Section 5 shows that their Gray images
are optimal when they satisfy certain conditions, and describes the application of image codes to secret sharing schemes. Section 6 puts the obtained results into perspective, and makes some conjectures for future research.
\section{Preliminaries}
\subsection{Rings}
\hspace*{0.5cm}We consider the ring $\mathbb{F}_2+v\mathbb{F}_2+v^2\mathbb{F}_2$ denoted by $R$, with $v^3=1$.
The ring $R$ has two maximal ideals, namely, $I_{1+v}=\{(1+v)(a_0+a_1v+a_2v^2):a_i\in\mathbb{F}_2,i=0,1,2\}$ and $I_{1+v+v^2}=\{0,1+v+v^2\}$.
Thus it is a semi-local, principal ideal ring. Given a positive integer $m$, we can construct the ring extension $R_m=\mathbb{F}_{2^m}+v\mathbb{F}_{2^m}+v^2\mathbb{F}_{2^m}$, with $v^3=1$. Let $\omega\in\mathbb{F}_4$ be a root of the irreducible polynomial $1+x+x^2\in\mathbb{F}_2[x]$. By a simple calculation, we get the decomposition of $v^3-1$ as follow:
$v^3-1=(v+1)(1+v+v^2)$ in $\mathbb{F}_2$ and $v^3-1=(v+1)(v+\omega)(v+\omega^2)$ in $\mathbb{F}_4=\{0,1,\omega,\omega^2\}$. Then the ring $R_m$ is isomorphic to $\mathbb{F}_{2^m}\bigoplus\mathbb{F}_{4^m}$ when $m$ is odd, and $\mathbb{F}_{2^m}\bigoplus\mathbb{F}_{2^m}\bigoplus\mathbb{F}_{2^m}$ when $m$ is even. Here $R^*_m$ denotes the group of units in $R_m$, and $\mathbb{F}^*_{2^m}$ denotes the multiplicative cyclic group of nonzero elements of $\mathbb{F}_{2^m}$. Taking account of the Chinese Remainder Theorem, we have ${R}_m^*\cong\mathbb{F}_{2^m}^*\oplus\mathbb{F}_{4^m}^*$ when $m$ is odd, and ${R}_m^*\cong\mathbb{F}_{2^m}^*\oplus\mathbb{F}_{2^m}^*\oplus\mathbb{F}_{2^m}^*$ when $m$ is even.\\
\hspace*{0.6cm}There is a Frobenius operator $F$ which maps $\alpha+\beta v+\gamma v^2$ onto $\alpha ^2+\beta ^2v+\gamma ^2v^2$. The \emph{Trace} function, denoted by $Tr$ is then defined as
$$Tr=\sum_{j=0}^{m-1}F^j.$$
\hspace*{0.5cm}It is immediate to check that
$Tr(\alpha+\beta v+\gamma v^2)=tr(\alpha)+tr(\beta)v+tr(\gamma)v^2,$
 for all $\alpha,\beta,\gamma\in \mathbb{F}_{2^m}$.
Here $tr()$ denotes the standard trace of $\mathbb{F}_{2^m}$ over $\mathbb{F}_2$. The following lemma describes a property of the trace function $Tr()$.\\
\noindent{\bf Lemma 2.1}
 If for all $x\in {R}_m,$ we have that $Tr(ax)=0,$ then $a=0.$ \\
{\bf Proof.} Let $a=a_1+a_2v+a_3v^2$ and $x=x_1+x_2v+x_3v^2$, where $a$ is a fixed element of $R$ and $a_i,x_i \in \mathbb{F}_{2^m}, i=1,2,3.$ Let $A_1=a_1x_1+a_2x_3+a_3x_2, A_2=a_1x_2+a_2x_1+a_3x_3, A_3=a_1x_3+a_2x_2+a_3x_1$, then $ax=(a_1+a_2v+a_3v^2)(x_1+x_2v+x_3v^2)=A_1+A_2v+A_3v^2.$
 Thus $Tr(ax)=0$ is equivalent to $tr(A_j)=0,j=1,2,3$. The $tr()$ we considered here is nondegenerate, and thus $tr(A_j)=0$. In consideration of the arbitrariness of $x_i$, we can obtain $a_i=0$, $i=1,2,3$. Thus $a=0$. This completes the proof. \qed
\subsection{Gray map}
\noindent\textbf{Definition 2.2} The Gray map $\phi$ from $R$ to $\mathbb{F}_2^3$ is defined by $$\phi(a_1+a_2v+a_3v^2)=(a_1,a_2,a_3),$$
where $a_1, a_2, a_3\in\mathbb{F}_2$, $i=1,2,3$.\\
\hspace*{0.5cm}It is a one to one map from ${{R}}$ to ${\mathbb{F}^3_2}$, which can be extended naturally to a map from ${{R}}^n$ to ${\mathbb{F}_2^{3n}}$. The Lee weight of $\alpha+\beta v+\gamma v^2$ is defined as the Hamming weight of its Gray image, namely, $w_L(\alpha+\beta v+\gamma v^2)=w_H(\alpha)+w_H(\beta)+w_H(\gamma)$, for $\alpha,\beta,\gamma\in{\mathbb{F}^n_2}$. The Lee distance of $x,y\in {{R}^n}$ is defined as $w_L(x-y)$. Thus the Gray map is a linear isometry from $({{R^n}},d_L)$ to $(\mathbb{F}^{3n}_2,d_H)$, where $d_L$ and $d_H$ denote the Lee and Hamming distance in $R^n$ and $\mathbb{F}^{3n}_2$, respectively.
\subsection{Codes}
\hspace*{0.5cm}A {\em linear} code C over ${{R}}$ of length $n$ is an $R$-submodule of $R^n$. The {\em inner product} of $x=(x_1,x_2,\ldots,x_n)$ and $y=(y_1,y_2,\ldots,y_n)\in{R^n}$ is defined by $\langle x,y\rangle=\sum\limits^n_{i=1}x_iy_i$, where the operation is performed in ${{R}}$. The dual code of $C$ is denoted by $C^\perp$ and defined as $C^\perp=\{y\in{{R}}^n\mid\langle x,y\rangle=0,\forall x\in C\}$. $C^\perp$ is also a linear code over ${{R}}$.\\
\hspace*{0.5cm}Given a finite abelian group $G,$ a code over $R$ is said to be {\bf abelian} if it is an ideal of the group ring $R[G].$ In other words, the coordinates of $C$ are indexed by elements of $G$ and $G$ acts regularly on this set. In the special case when $G$ is cyclic, the code is a cyclic code in the usual sense \cite{MJ}.\\
\noindent\textbf{Definition 2.3} For $a\in {R}_m$, we define the vector $Ev(a)$ by the following evaluation map: $$Ev(a)=(Tr(ax))_{x\in {R}_m^*}.$$ Define the code $C_m$ by the formula $C_m=\{Ev(a)|a\in {R}_m\}$. Thus $C_m$ is a code of length $|{R}_m^*|$ over $R$.

\section{Statement of main results}
\hspace*{0.5cm}We are now in a position to state the main results of this paper. First, we give the Lee weight distribution of our codes.\\
\noindent\textbf{Theorem 3.1}\label{enum} For $a\in {R}_m, m$ is odd, the Lee weight distribution of the codewords of $C_m$ is given below.
\begin{enumerate}
{\item[(i)] If $a=0$, then $w_{L}(Ev(a))=0$;
\item[(ii)] If $a\in {R}_m^*$, then $w_{L}(Ev(a))=3(2^{3m-1}-2^{2m-1}-2^{m-1})$;
\item[(iii)] If $a\in {{R}_m}\backslash\{{R}_m^*\cup\{0\}\}$, $a=a_1+a_2v+a_3v^2$,\\
1) if $(a_1+a_2,a_1+a_3)=(0,0)$, $a_1+a_2+a_3\neq0$, then $w_{L}(Ev(a))=3(2^{3m-1}-2^{m-1})$;\\
2) if $(a_1+a_2,a_1+a_3)\neq(0,0)$, $a_1+a_2+a_3=0$, then $w_{L}(Ev(a))=3(2^{3m-1}-2^{2m-1})$.
}
 \end{enumerate}
\noindent\textbf{Theorem 3.2}\label{enum} For $a\in {R}_m, m$ is even, the Lee weight of the codewords of $C_m$ is given below.
\begin{enumerate}
{
\item[(i)] If $a=0$, then $w_{L}(Ev(a))=0$;
\item[(ii)] If $a\in {R}_m^*$, then $w_{L}(Ev(a))=3(2^{3m-1}-3\cdot2^{2m-1}+3\cdot2^{m-1})$;
\item[(iii)] If $a\in {{R}_m}\backslash\{{R}_m^*\cup\{0\}\}$, where $a=a_1+a_2v+a_3v^2$,\\
1) if one of three terms $a_1+a_2+a_3,a_1+a_2\omega+a_3\omega^2,a_1+a_2\omega^2+a_3\omega$ equals to $0$, \hspace*{0.4cm}then $w_{L}(Ev(a))=3(2^{3m-1}-3\cdot2^{2m-1}+2^m)$;\\
2) if one of three terms $a_1+a_2+a_3,a_1+a_2\omega+a_3\omega^2,a_1+a_2\omega^2+a_3\omega$ not equals to \hspace*{0.4cm}$0$, then $w_{L}(Ev(a))=3(2^{3m-1}-2^{2m}+2^{m-1})$.
}
 \end{enumerate}

 Next, the optimality of their binary images is given in the following theorem.

\noindent\textbf{Theorem 3.3}
 For any odd $m>1$, the code $\phi(C_m)$ is optimal for given length and dimension.

 Eventually, we study their dual Lee distance.

\noindent\textbf{Theorem 3.4}
 For all $m>1,$ the dual Lee distance $d'$ of $C_m$ is $2.$

It is well-known that linear codes have application in secret sharing scheme (SSS). To determine the set of all minimal access of an SSS,
the concept of minimal codewords for a partial order on the codewords
was introduced. The {\em support} $s(x)$ of a vector $x$ in $\mathbb{F}_q^n$ is defined as the set of indices where it is nonzero.
We say that a vector $x$ {\em covers} a vector $y$ if $s(x)$ contains $s(y)$. \\
\noindent\textbf{Definition 3.5} A {\em minimal} codeword of a linear code $C$ is a nonzero codeword that does not cover any other nonzero codeword.\\
\noindent\textbf{Proposition 3.6}
 All the nonzero codewords of $\phi(C_m),$ for $m> 1$ and $m$ is odd, are minimal.\\
\noindent\textbf{Proposition 3.7}
 All the nonzero codewords of $\phi(C_m),$ for $m>0$ and $m$ is even, are minimal.
\section{Proof of Theorems 3.1, 3.2, 3.3 and 3.4}
\hspace*{0.5cm}First, we give some auxiliary lemmas, which will be employed in the proof of the main theorems.\\
\noindent\textbf{Lemma 4.1}\cite[Lemma 4.1]{SL} If $\mathbf{y}=(y_1,y_2,\ldots,y_n)\in \mathbb{F}_2^n,$
then $2w_H(\mathbf{y})=n-\sum\limits_{i=1}^n(-1)^{y_i}.$\\
\noindent\textbf{Lemma 4.2}\cite[Lemma 9 p.143]{MJ} If $z \in \mathbb{F}_{2^m}^*,$ then $\sum\limits_{x\in \mathbb{F}_{2^m}}(-1)^{tr(z x)}=0.$\\
\hspace*{0.5cm}If $x=a+bv+cv^2\in R^n$, where $a,b,c\in \mathbb{F}_2^n$, in view of Definition 2.2, we get $w_L(x)=w_H(a)+w_H(b)+w_H(c)$. By using Lemma 4.1, we have $3n-2w_L(x)=\sum\limits^n_{i=1}(-1)^{a_i}+\sum\limits^n_{i=1}(-1)^{b_i}+\sum\limits^n_{i=1}(-1)^{c_i}$, where $a=(a_1,a_2,\ldots,a_n)$, $b=(b_1,b_2,\ldots,b_n)$ and $c=(c_1,c_2,\ldots,c_n)$.\\

Suppose $m$ is odd, according to the Chinese Remainder Theorem, we can decompose $r+sv+tv^2$ in $R_m$, $r+sv+tv^2=(v^2+v+1)\theta_1+(v+1)(\theta_2+\theta_3v)$, where $\theta_1,\theta_2,\theta_3\in \mathbb{F}_{2^m}$.
Comparing coefficients, we have the following system of equations:
\begin{equation*}
 \begin{cases}
\theta_1+\theta_2=r\\
\theta_1+\theta_2+\theta_3=s\\
\theta_1+\theta_3=t\\
\end{cases},~~~~~~~\begin{cases}
\theta_1=s+r+t\\
\theta_2=s+t\\
\theta_3=s+r\\
\end{cases},
\end{equation*}
which implies ${R}_m^*=\{x_1+x_2v+x_3v^2:x_1+x_2+x_3\neq0,(x_1+x_2,x_1+x_3)\neq(0,0),x_1,x_2,x_3\in\mathbb{F}_{2^m}\}$ and $|R_m^*|=(2^m-1)(2^{2m}-1).$\\
\hspace*{0.5cm}For convenience, we adopt the following notations unless otherwise stated in this section. Let $I_1=\sum\limits_{i=1}^3x_i$, $I_2=x_1+x_2$ and $I_3=x_1+x_3$. Then $R_m^*=\Big\{\sum\limits_{i=1}^3x_iv^i:I_1\neq0,(I_2,I_3)\neq(0,0),I_1,I_2,I_3\in\mathbb{F}_{2^m}\Big\}.$ The proof of Theorem 3.1 is given below.
\subsection{Proof of Theorem 3.1}
{\bf Proof.}
Let $x=x_1+x_2v+x_3v^2\in {R^*_m}$, where $I_1\neq0$, $(I_2,I_3)\neq(0,0)$.\\
\hspace*{0.5cm}(i) If $a=0$, then $Ev(a)=(0,0,\ldots,0)$. So $w_L(Ev(a))=0$.\\
\hspace*{0.5cm}(ii) If $a=a_1+a_2v+a_3v^2\in{R}_m^*$, then $a_1+a_2+a_3\neq0$ and $(a_1+a_2,a_1+a_3)\neq(0,0)$. Let $A=a_1+a_2+a_3$, $B=a_1+a_2$, $C=a_1+a_3$,
then we can simplify $ax$ as follows:
\begin{eqnarray*}
                         ax &=& (a_1x_1+a_2x_3+a_3x_2)+(a_1x_2+a_2x_1+a_3x_3)v+(a_1x_3+a_2x_2+a_3x_1)v^2\\
                            &=& I_1(a_1+a_2+a_3)+I_2(a_1+a_2)+I_3(a_1+a_3)+(I_1(a_1+a_2+a_3)+I_2(a_2+a_3)+\\
                            && I_3(a_1+a_2))v
                            +(I_1(a_1+a_2+a_3)+I_2(a_1+a_3)+I_3(a_2+a_3))v^2\\
                            &=& I_1 A+I_2 B+I_3 C+(I_1 A+I_2 (B+C)+I_3 B)v+(I_1 A+I_2 B+I_3 (B+C))v^2.
                                                  \end{eqnarray*}
In terms of Definition 2.3, we have
\small\begin{eqnarray*}
                         \phi(Ev(a)) &=& \phi({(Tr(ax))}_{x\in {R_m^*}})\\
                          &=&({tr(I_1 A+I_2 B+I_3 C)},{tr(I_1 A+I_2 (B+C)+I_3 B)},{tr(I_1 A+I_2 B+I_3 (B+C))}).
                        \end{eqnarray*}
In the light of Lemmas 4.1 and 4.2, we can obtain
\begin{eqnarray*}
                         3|R^*_m|-2w_L(Ev(a)) &=& \sum\limits_{{I_1\neq0},{(I_2,I_3)\neq(0,0)}}(-1)^{tr(I_1 A+I_2 B+I_3 C)}
                         \\&&+\sum\limits_{{I_1\neq0},{(I_2,I_3)\neq(0,0)}}(-1)^{tr(I_1 A+I_2 (B+C)+I_3 B)}\\&&+\sum\limits_{{I_1\neq0},{(I_2,I_3)\neq(0,0)}}(-1)^{tr(I_1 A+I_2 B+I_3 (B+C))}\\
                          &=&3,
                        \end{eqnarray*}
which implies $w_L(Ev(a))=3(2^{3m-1}-2^{2m-1}-2^{m-1}).$\\
\hspace*{0.5cm}(iii) Let $a=a_1+a_2v+a_3v^2\in {{R}_m}\backslash\{{R}_m^*\cup\{0\}\}$.\\
\hspace*{0.5cm}1) If $(a_1+a_2,a_1+a_3)=(0,0)$, $a_1+a_2+a_3\neq0$, then we claim $a=a_1(1+v+v^2)$, where $a_1=0$. By a simple calculation, we can obtain $ax=a_1I_1+a_1I_1v+a_1I_1v^2$.
From Definition 2.3, we have
$$\phi(Ev(a))=({tr(a_1I_1)},{tr(a_1I_1)},{tr(a_1I_1)}).$$
Applying Lemmas 4.1 and 4.2, yields
$$ 3|R^*_m|-2w_L(Ev(a))=3\sum_{{I_1\neq0},{(I_2,I_3)\neq(0,0)}}(-1)^{tr(a_1I_1)}=3(1-2^{2m}).$$
So $w_L(Ev(a))=3(2^{3m-1}-2^{m-1})$.\\
\hspace*{0.5cm}2) If $(a_1+a_2,a_1+a_3)\neq(0,0)$, $a_1+a_2+a_3=0$, then
$ax=((a_1+a_2)I_2+a_2I_3)+((a_1+a_2)I_3+a_1I_2)v+(a_1I_3+a_2I_2)v^2.$
In the light of Definition 2.3, we have
\begin{eqnarray*}
                       \phi(Ev(a))&=& ({tr((a_1+a_2)I_2+a_2I_3)},{tr(a_1I_2+(a_1+a_2)I_3)},{tr(a_2I_2+a_1I_3)})\\
                       &=:& (\alpha_1,\alpha_2,\alpha_3).
                        \end{eqnarray*}
Armed with Lemmas 4.1 and 4.2, we get
\begin{eqnarray*}
3|R^*_m|-2w_L(Ev(a))  &=& \sum\limits_{{I_1\neq0},{(I_2,I_3)\neq(0,0)}}(-1)^{\alpha_1}
 +\sum\limits_{{I_1\neq0},{(I_2,I_3)\neq(0,0)}}(-1)^{\alpha_2}\\
 &&+\sum\limits_{{I_1\neq0},{(I_2,I_3)\neq(0,0)}}(-1)^{\alpha_3}\\
                             &=& 3(1-2^m).
                        \end{eqnarray*}
It is immediate to obtain that $w_{L}(Ev(a))=3(2^{3m-1}-2^{2m-1})$.\qed\\

Combining with Theorem 4.3 and Lemma 2.1, we have constructed a binary linear code of length $3|R^*_m|$, of dimension $3m$, with three nonzero weights $w_1,w_2$ and $w_3$, with frequencies $f_1,f_2$ and $f_3$, respectively.
We list the values of these parameters in Table I.
\begin{center}$\mathrm{Table~I. }$~~~$\mathrm{weight~ distribution~ of}~ C_m~(m~ \mathrm{is~ odd})$\
\begin{tabular}{ccc||cc}
\hline
  Weight& &&& Frequency  \\
  \hline
  0        & &&& 1\\
  \ \ \ \ \ \ \ \ \ \ \ \ $w_1=3(2^{3m-1}-2^{2m-1}-2^{m-1})$       &  &&&              \ \ \ \ \ \ \ \ \ \ \ \ \ \ \ $f_1=(2^m-1)(2^{2m}-1)$\\
  $w_2=3(2^{3m-1}-2^{m-1})$                                                  &  &&&              $f_2=2^m-1$\\
  \ $w_3=3(2^{3m-1}-2^{2m-1})$         &  &&& \ $f_3=2^{2m}-1$\\
  \hline
\end{tabular}
\end{center}

Suppose $m$ is even, we can obtain the decomposition of $r+sv+tv^2$ in $R_m$, from the Chinese Remainder Theorem. That is $r+sv+tv^2=a_0(v-\omega)(v-\omega^2)+a_1(v-1)(v-\omega^2)+a_2(v-1)(v-\omega)$, where $a_0,a_1,a_2\in \mathbb{F}_{2^m}$.
Comparing coefficients, we have
\begin{equation*}
 \begin{cases}
a_0+a_1\omega^2+a_2=r\\
a_0+a_1\omega+a_2\omega^2=s\\
a_0+a_1+a_2=t\\
\end{cases},~~~~~~~\begin{cases}
a_0=r+s+t\\
a_1=\frac{r+\omega s+\omega^2t}{\omega^2}\\
a_2=\frac{r+\omega^2s+\omega t}{\omega}\\
\end{cases}.
\end{equation*}
Then we can obtain
${R}_m^*=\{x_1+x_2v+x_3v^2:x_1+x_2+x_3\neq0,x_1+x_2\omega +x_3\omega^2\neq0,x_1+x_2\omega^2+x_3\omega\neq0, x_1, x_2, x_3\in\mathbb{F}_{2^{m}}\}$ and $|{R}_m^*|=(2^m-1)^3.$\\
\hspace*{0.5cm}For convenience, we adopt the following notations unless otherwise stated in this subsection. Let $I_4=\sum\limits_{i=1}^3x_i$, $I_5=x_1+x_2\omega+x_3\omega^2$ and $I_6=x_1+x^2\omega^2+x_3\omega$, then $R_m^*=\Big\{\sum\limits_{i=1}^3x_iv^i:I_4,I_5,I_6\in\mathbb{F}^*_{2^m}\Big\}.$
Now we turn to the proof of Theorem 3.2.
\subsection{Proof of Theorem 3.2}
{\bf Proof.}
(i) If $a=0$, then $Ev(a)=(0,0,\ldots,0)$. So $w_L(Ev(a))=0.$\\
\hspace*{0.5cm}(ii) If $a=a_1+a_2v+a_3v^2, x=x_1+x_2v+x_3v^2\in R^*_m$, then $a_1+a_2+a_3\neq0, a_1+a_2\omega+a_3\omega^2\neq0, a_1+a_2\omega^2+a_3\omega\neq0$, $I_i\in\mathbb{F}_{2^m}^*, i=4,5,6$. Let $D=a_1+a_2+a_3$, $E=a_1+a_2\omega+a_3\omega^2$ and $F=a_1+a_2\omega^2+a_3\omega$. We can simplify $ax$ as follows:
\begin{eqnarray*}
                         ax &=& (a_1x_1+a_2x_3+a_3x_2)+(a_1x_2+a_2x_1+a_3x_3)v+(a_1x_3+a_2x_2+a_3x_1)v^2\\
                          &=&DI_4+EI_5+FI_6+(DI_4+FI_5+EI_6)v+(DI_4+\omega^2 EI_5+\omega FI_6)v^2.
                        \end{eqnarray*}
We deduce from Definition 2.3 that
\small\begin{eqnarray*}
                         \phi(Ev(a))&=&\phi({(Tr(ax))}_{x\in{R_m^*}})\\
                          &=&(tr(DI_4+EI_5+FI_6),tr(DI_4+FI_5+EI_6),tr(DI_4+\omega^2 EI_5+\omega FI_6)).
                        \end{eqnarray*}
By a simple calculation, we have
\small\begin{eqnarray*}
                         3|R^*_m|-2w_L(Ev(a))&=&\sum\limits_{I_4,I_5,I_6\in\mathbb{F}_{2^m}^*}(-1)^{tr(DI_4+EI_5+FI_6)}
+\sum\limits_{I_4,I_5,I_6\in\mathbb{F}_{2^m}^*}(-1)^{tr(DI_4+FI_5+EI_6)}
\\&&+\sum\limits_{I_4,I_5,I_6\in\mathbb{F}_{2^m}^*}(-1)^{tr(DI_4+\omega^2 EI_5+\omega FI_6)}\\
                          &=&-3.
                        \end{eqnarray*}
Therefore, $w_{L}(Ev(a))=3(2^{3m-1}-3\cdot2^{2m-1}+3\cdot2^{m-1})$.\\
\hspace*{0.5cm}(iii) Set $a=a_1+a_2v+a_3v^2\in {{R}_m}\backslash\{{R}_m^*\cup\{0\}\}$.\\
\hspace*{0.5cm}1) If one of three terms $D,E$ and $F$ equals $0$. Without loss of generality, we can let $D=0, E\neq0$ and $F\neq0$. On the same principle, we have
$\phi(Ev(a))=(DI_4+EI_5+FI_6,DI_4+FI_5+EI_6,DI_4+\omega^2 EI_5+\omega FI_6).$
Since $D=0, E\neq0$ and $F\neq0$, by a simple calculation, we can obtain
\small\begin{eqnarray*}
                         3|R^*_m|-2w_L(Ev(a))&=&\sum\limits_{I_4,I_5,I_6\in\mathbb{F}_{2^m}^*}(-1)^{tr(DI_4+EI_5+FI_6)}
+\sum\limits_{I_4,I_5,I_6\in\mathbb{F}_{2^m}^*}(-1)^{tr(DI_4+FI_5+EI_6)}\\
&&+\sum\limits_{I_4,I_5,I_6\in\mathbb{F}_{2^m}^*}(-1)^{tr(DI_4+\omega^2 EI_5+\omega FI_6)}\\
                          &=&3(2^m-1).
                        \end{eqnarray*}
Then $w_{L}(Ev(a))=3(2^{3m-1}-3\cdot2^{2m-1}+2^m)$.\\
\hspace*{0.5cm}2) If one of three terms $D,E$ and $F$ not equals $0$. Without loss of generality, we can let $D=0, E=0$ and $F\neq0$. We can use a similar approach as 1) to obtain $3|R^*_m|-2w_L(Ev(a))=3(-2^{2m}+2^{m+1}-1)$, and omit it here. Furthermore, it is clear to know $w_{L}(Ev(a))=3(2^{3m-1}-2^{2m}+2^{m-1}).$\qed\\

In the light of Theorem 3.2 and Lemma 2.1, we have constructed a binary linear code of length $3|R^*_m|$, of dimension $3m$, with three nonzero weights $w_1,w_2$ and $w_3$, with frequencies $f_1,f_2$ and $f_3$, respectively.
In Table II, we present the values of these parameters.
\begin{center}$\mathrm{Table~II. }$~~~$\mathrm{weight~ distribution~ of}~ C_m~(m~ \mathrm{is~ even})$\
\begin{tabular}{ccc||cc}
\hline
  Weight& &&& Frequency  \\
  \hline
  0        & &&& 1\\
  \ \ \ \ \ \ \ \ \ \ $w_1=3(2^{3m-1}-3\cdot2^{2m-1}+3\cdot2^{m-1})$       &  &&&              \ \ \ \ $f_1=(2^m-1)^3$\\
  \ \ \ $w_2=3(2^{3m-1}-3\cdot2^{2m-1}+2^m)$                                                  &  &&&              \ \ \ \ $f_2=(2^m-1)^2$\\
  $w_3=3(2^{3m-1}-2^{2m}+2^{m-1})$         &  &&& $f_3=2^{m}-1$\\
  \hline
\end{tabular}
\end{center}

\noindent\textbf{Remark} No matter $m$ is odd in Table I or even in Table II, $\phi(C_m)$ is a binary three-weight linear code and its parameters are different from those in \cite{CG,DD,HY} and \cite{SL}, which implies that the obtained codes in our paper are new. The parameters of $\phi(C_m)$ depend on the value of $m$. Next, we give two examples to illustrate it.\\
\noindent\textbf{Example 4.5}
 Let $m=3$. Since $m$ is odd, by Table I, we can obtain a three-weight binary linear code of length $1323$, of dimension $9$, with three nonzero weights $660$, $756$ and $672$, and frequencies $441$, $7$ and $63$, respectively.\\
\noindent\textbf{Example 4.6}
 Let $m=2$. Since $m$ is even, by Table II, we can obtain a three-weight binary linear code of length $81$, of dimension $6$, with three nonzero weights $42$, $36$ and $54$, and frequencies $27$, $9$ and $3$, respectively.
\subsection{Proof of Theorem 3.3}
\hspace*{0.5cm}The proof of Theorem 3.3 is presented as follows:\\
{\bf Proof.}
If $m\geq3$ is odd, $|{R}_m^*|=2^{3m}-2^{2m}-2^m+1$.
Recall that the parameters of $\phi(C_m)$ are $[n,K,d]=[3|R_m^*|,3m,d],$ where $d=3(2^{3m-1}-2^{2m-1}-2^{m-1})$. We claim that
$\sum\limits_{i=0}^{K-1}\lceil \frac{d+1}{2^i} \rceil > n,$ violating the Griesmer bound. Indeed, depending on the range of $i$, five expressions for the inner ceiling function may occur,
\begin{itemize}
\item If $0\leq i\leq m-1,$ then $\lceil \frac{d+1}{2^i} \rceil = 3\cdot(2^{3m-1-i}-2^{2m-1-i}-2^{m-1-i})+1$;
\item If $i=m$, then $\lceil \frac{d+1}{2^i} \rceil = 3\cdot(2^{2m-1}-2^{m-1})-1$;
\item If $m+1\leq i\leq 2m-1,$ then $\lceil \frac{d+1}{2^i} \rceil =3\cdot(2^{3m-1-i}-2^{2m-1-i})$;
\item If $i=2m$, then $\lceil \frac{d+1}{2^i} \rceil =3\cdot2^{m-1}-1$;
\item If $2m+1\leq i\leq3m-1,$ then $\lceil \frac{d+1}{2^i} \rceil =3\cdot2^{3m-1-i}$.
\end{itemize}
Thus \begin{eqnarray*}
       \sum_{i=0}^{3m-1}\Big \lceil \frac{d+1}{2^i} \Big \rceil
       &=& \sum_{i=0}^{m-1}\Big\lceil \frac{d+1}{2^i}\Big\rceil+3(2^{2m-1}-2^{m-1})-1+\sum_{i=m+1}^{2m-1}\Big\lceil \frac{d+1}{2^i}\Big\rceil+3\cdot2^{m-1}-1
       \\&&+\sum_{i=2m+1}^{3m-1}\Big\lceil \frac{d+1}{2^i}\Big\rceil \\
        &=&3\cdot(2^{3m}-2\cdot2^{2m}+1)+m+3\cdot(2^{2m-1}-2^{m-1})-1+3\cdot(2^{2m-1}-3\cdot2^{m-1}\\
        &&+1)+3\cdot2^{m-1}-1+3\cdot(2^{m-1}-1)\\
        &=& 3(2^{3m}-2^{2m}-2^{m})+1+m.
     \end{eqnarray*}
When $m\geq3$, $\sum_{i=0}^{3m-1}\lceil \frac{d+1}{2^i} \rceil=3(2^{3m}-2^{2m}-2^{m})+1+m>3(2^{3m}-2^{2m}-2^{m}+1)=3|R_m^*|$. This completes the proof.\qed\\
\noindent{\bf Example 4.7} Let $C_3$ be the code considered in Example 4.5, we can verify that $C_3$ is optimal.
\subsection{Proof of Theorem 3.4}

\hspace*{0.5cm}We investigate the dual Lee distance of $C_m$ in this section. As before we study another property of the trace function. The proof of the following lemma is similar to Lemma 2.1, so we omit the details here.\\
\noindent{\bf Lemma 4.8}
 If for all $a\in {R}_m,$ we have that $Tr(ax)=0,$ then $x=0.$

Armed with Lemma 4.8, we can prove Theorem 3.4 as follows: It is enough to prove that $C_m^\perp$ does not contains a codeword with Lee weight $1$, but contains a codeword with Lee weight $2$.

If $C_m^\perp$ contains a codeword of Lee weight $1$, then its nonzero digit has values $v^j$, $j=0,1,2$, where $v^0=1$. This implies that $\forall a\in {R}_m$, $v^jTr(ax)=0$ at some $x\in{R}_m^*$. Let $a=a_1+a_2v+a_3v^2$ and $x=x_1+x_2v+x_3v^2$.
Using Lemma 4.8, we conclude, by the nondegenerate character of $tr()$, that $x_i=0.$ Contradiction with $x=x_1+x_2v+x_3v^2\neq0$.

 If we can find a codeword of $C_m^\perp$ which has two digits $1$ and $v$ at some $x,y\in{R^*_m}$, where $x=x_1+x_2v+x_3v^2$ and $y=y_1+y_2v+y_2v^2$, then we prove $d^\prime=2$. If such codeword exists, then $\forall a\in {R}_m$, we have $1\cdot Tr(ax)+v\cdot Tr(ay)=0$, where $x,y\in{R_m^*}.$ By a direct calculation we get
\small\begin{eqnarray*}
  Tr(ax)+v\cdot Tr(ay)
  &=& tr(a_1x_1+a_2x_3+a_3x_2+a_1y_3+a_2y_2+a_3y_1)+v(tr(a_1x_2+a_2x_1+a_3x_3\\
  &&+a_1y_1+a_2y_3+a_3y_2))+v^2(tr(a_1x_3+a_2x_2+a_3x_1+a_1y_2+a_2y_1+a_3y_3))\\
  &=:& tr(A_4)+vtr(A_5)+v^2tr(A_6).
\end{eqnarray*}
Using the nondegenerate character of $tr()$, we have $tr(A_j)=0, j=4,5,6.$ In consideration of the arbitrariness of $a_i,i=1,2,3$, we can obtain
$x_1=y_3$, $x_2=y_1$ and $x_3=y_2$.
Then there exists a codeword of $C_m^\perp$ such that its Lee weight is $2$, which implies $d^\prime=2$.
\section{Proof of Propositions 3.6 and 3.7 and secret sharing schemes}
\subsection{The sufficient condition of minimal codeword}
\hspace*{0.5cm}Minimal codeword in $q$-ary linear codes arise in numerous applications, for instance, in constructing decoding algorithms and studying SSS. In general, the problem of determining the minimal codewords of a given $q$-ary linear code is difficult. But when the weights of a given linear code $C$ are close enough to each other, then each nonzero codeword of $C$ is minimal, as described by the following lemma \cite{AB}.\\
\noindent{\bf Lemma 5.1 }(Ashikhmin-Barg) Denote by $w_0$ and $w_{\infty}$ the minimum and maximum nonzero weights of a $q$-ary code $C$, respectively. If
$$\frac{w_0}{w_{\infty}}>\frac{q-1}{q},$$ then every nonzero codeword of $C$ is minimal.
\subsection{Proof of Proposition 3.6}
{\bf Proof.}
 We apply Lemma 5.1 with $w_0=w_1=3(2^{3m-1}-2^{2m-1}-2^{m-1})$ and $w_{\infty}=w_2=3(2^{3m-1}-2^{m-1})$ as per Table I. Rewriting the inequality of the lemma as $2w_0>w_{\infty},$ we obtain successively
 $$2w_0-w_{\infty}=3(2^{3m-1}-2^{2m}-2^{m-1})>0.$$
Hence the proposition is proved.\qed
\subsection{Proof of Proposition 3.7}
{\bf Proof.}
 We apply Lemma 5.1 with $w_0=w_2=3(2^{3m-1}-3\cdot2^{2m-1}+2^{m}),$ and $w_{\infty}=w_3=3(2^{3m-1}-2^{2m}+2^{m-1})$ as per Table II. Rewriting the inequality of the lemma as $2w_0>w_{\infty},$ we obtain successively
 $$2w_0-w_{\infty}=3(2^{3m-1}-2^{2m}+3\cdot2^{m-1})>0.$$
Hence the proposition is proved.\qed
\subsection{Secret sharing schemes}
\hspace*{0.5cm} When all nonzero codewords are minimal, it was shown in \cite{YC} that there is the following alternative, depending on $d'$:
\begin{itemize}
 \item If $d'\ge 3,$ then the SSS is \emph{``democratic''}: every user belongs to the same number of coalitions.
 \item If $d'=2,$  then there are users  who belong to every coalition: the \emph{``dictators''.}
\end{itemize}
\hspace*{0.5cm}Depending on the application, one or the other situation might be more suitable. By Theorem 3.4 and Propositions 3.6 and 3.7, we see that two secret sharing schemes built on $\phi(C_m)$ are dictatorial.


\section{Conclusion}
\hspace*{0.6cm}In this paper, we have studied a family of trace codes over $\mathbb{F}_2+v\mathbb{F}_2+v^2\mathbb{F}_2.$
These codes are provably abelian, but not visibly cyclic. Using a character sum approach, we have been able to describe their weight distributions, and
we have obtained a class of abelian binary three-weight codes by application of the Gray map. These latter codes are shown to be optimal under some conditions on the parameter $m$
by considering the Griesmer bound. Moreover, the parameters of the obtained codes in Table I and Table II are different from those in \cite{CG,DD,HY} and \cite{SL}, which implies that the obtained codes in our paper are new. It is worth exploring more general constructions by varying the alphabet of the code, the Gray map, or the localizing set of the trace code.





\begin{thebibliography}{99}\addtolength{\itemsep}{-7pt}
\bibitem{AB} Ashikhmin, A., Barg, A.: Minimal vectors in linear codes, IEEE Transactions on Information Theory, 1998, 44(5):2010-2017.

\bibitem{B} Bonnecaze, A., Bracco, A., Dougherty, S., Nochefranca, L., Sol\'e, P.: Cubic self-dual binary codes, IEEE Transactions on Information Theory, 2003, 49(9):2253-2258.
\bibitem{AE} Brouwer, A., Haemers, W.: {\it Spectra of Graphs}, Springer New York, 2012.

\bibitem{CG} Calderbank, A., Goethals, J.: Three-weight codes and association schemes, Philips Journal of Research, 1984, 39(4):143-152.

\bibitem{RWM}Calderbank, A., Kantor, W.: The geometry of two-weight codes, Bulletin of the London Mathematical Society, 1986, 18(2):97-122.

\bibitem{DP} Delsarte, P.: Weights of linear codes and strongly regular normed spaces, Discrete Mathematics, 1972, 3(1-3):47-64.

\bibitem{DD} Ding, C., Ding, K.: A class of two-weight and three-weight codes and their applications in secret sharing, IEEE Transactions on Information Theory, 2015, 61(11):5835-5842.

\bibitem{YC} Ding, C., Yuan, J.: Covering and secret sharing with linear codes, Lecture Notes in Computer Science, 2003, 2731:11-25.

\bibitem{DY} Ding, C., Yang, J.: Hamming weights in irreducible cyclic codes, Discrete Mathematics, 2011, 313(4):434-446.

\bibitem{DZ} Ding, C., Zhou, Z.: A class of the three-weight cyclic codes, IEEE Transactions on Communications, 2013, 25(10):79-93.

\bibitem{GG}Griesmer, J.: A Bound for Error-Correcting Codes, IBM Journal of Research \& Development, 1960, 4(5):532-542.

\bibitem{HY} Heng, Z., Yue, Q.: A class of binary linear codes with at most three weights, IEEE Communications Letters, 2015, 19(9):1488-1491.

 \bibitem{LS} Ling, S., Sol\'e, P.: On the algebraic structure of quasi-cyclic codes. I. Finite fields, IEEE Transactions on Information Theory, 2001, 47(7):2751-2760.
\bibitem{MJ} Macwilliams, F., Sloane, N.: {\it The theory of error-correcting codes}, 1977.

\bibitem{SL} Shi, M., Liu, Y., Sol\'{e}, P.: Optimal two weight codes over $\mathbb{F}_2+u\mathbb{F}_2$, IEEE Communications Letters, 2016, 20(12): 2346-
2349. 

\bibitem{SP} Shi, M., Liu, Y., Sol\'{e}, P.: Optimal two weight codes from trace codes over a non-chain ring, Discrete Applied Mathmatics, 219(2017)176-
181.

\bibitem{SR} Shi, M., Wu, R., Liu, Y., Sol\'{e}, P.: Two and three-weight codes over $\mathbb{F}_p+u\mathbb{F}_p$, Cryptography and Communications-Discrete Structures, 2017, 9(5): 637-646.

\bibitem{SS} Shi, M., Zhu, S., Yang, S.: A class of optimal $p$-ary codes from one-weight codes over $\F_p[u]/\langle u^m\rangle$, Journal of the Franklin Institute, 2013, 350(5):929-937.

\bibitem{YJ} Yuan, J., Ding, C.: Secret sharing schemes from three classes of linear codes, IEEE Transactions on Information Theory, 2006, 52(1):206-212.

\end{thebibliography}
\end{document}